\documentclass[12pt,notitlepage,a4paper]{article}

\usepackage[margin = 1 in]{geometry}
\usepackage{natbib,hyperref,amsmath,amssymb,graphicx,soul,setspace}
\hypersetup{
	colorlinks=true,
	linkcolor=red,
	citecolor=blue,
	urlcolor=blue,
	bookmarksnumbered=true,
	pdfstartview=
}

\def\sym#1{\ifmmode^{#1}\else\(^{#1}\)\fi}

\begin{document}

\title{Information extraction and artwork pricing}
\author{Jaehyuk Choi, Lan Ju, Jian Li, and Zhiyong Tu\footnote{Peking University HSBC Business School, University Town, Nanshan District, Shenzhen 518055, China, \url{jaehyuk@phbs.pku.edu.cn} (Choi), \url{julan@phbs.pku.edu.cn} (Ju), \url{lijian1991@pku.edu.cn} (Li), and \url{zytu@phbs.pku.edu.cn} (Tu). Corresponding author: Zhiyong Tu.}}
\maketitle
%\shortTitle{Art price and the Shannon information quantity}
%\author{Jaehyuk Choi, Lan Ju, Jian Li, and Zhiyong Tu\thanks{Peking University HSBC Business School, University Town, Nanshan District, Shenzhen 518055, China, \url{jaehyuk@phbs.pku.edu.cn} (Choi), \url{julan@phbs.pku.edu.cn} (Ju), \url{lijian1991@pku.edu.cn} (Li), and \url{zytu@phbs.pku.edu.cn} (Tu).}}
%\date{\today}
%\pubMonth{Month}
%\pubYear{Year}
%\pubVolume{Vol}
%\pubIssue{Issue}
%\JEL{G12, Z1, Z11}
%\Keywords{Art Pricing; Information; Entropy; Singular Value Decomposition}

\begin{abstract}

Traditional art pricing models often lack fine measurements of painting content. This paper proposes a new content measurement: the Shannon information quantity measured by the singular value decomposition (SVD) entropy of the painting image. Using a large sample of artworks' auction records and images, we show that the SVD entropy positively affects the sales price at 1\% significance level. Compared to the other commonly adopted content variables, the SVD entropy has advantages in variable significance, sample robustness as well as model fit. Considering the convenient availability of digital painting images and the straightforward calculation algorithm of this measurement, we expect its wide application in future research.

\end{abstract}
%\maketitle
% No introdution for JER
%\section{Introduction}

Artwork has become an increasingly important category in global investors' portfolios. During the past twenty years, the annual total amount of artwork auctioned worldwide has more than quadrupled, while the world GDP has just doubled.\footnote{Art Market Report 2020, Artmarket and AMMA.} For both art professionals and researchers, how to price artwork has always been a central concern. In literature, the most commonly used art pricing model is the hedonic model (e.g., \citet{buelens1993revisiting}, \citet{chanel1995art} and \citet{taylor2011price}). In this model, the sales price of a painting is often explained by multiple factors ranging from painting attributes such as size, material, and signature to sales conditions such as year, salesroom, and sales location.

These aforementioned price determinants include the physical features of the artwork as well as the economic conditions for the auction. However, the most important factor for the valuation, what the painting actually depicts, still lacks adequate consideration. We know that paintings' contents are extremely heterogeneous. To account for the content heterogeneity, researchers usually introduce such categorical variables as signature, dated, topic and style. The first two describe whether the painting is signed or dated by the artist, and the last two classify artworks into distinct subjects (topics) and movements (styles). For example,  \citet{renneboog2012buying} introduce eleven subject matters (e.g., abstract, animals, landscape) and thirteen movements (e.g., renaissance, baroque, pop) into the hedonic model. These content dummies undoubtedly improve the traditional art pricing model. Here, we ask whether we can capture deeper information unique to the painting that goes beyond these broad categorizations. After all, ample content variations exist for the both signed and dated, or within the sphere of the same subject and style.

Currently, many databases provide large-scale art data that include both auction records and high-quality digital images. Researchers have started applying computer graphics to analyze image data, so as to extract more pricing information from painting content. Recent research in this strand mainly focuses on the effect of the color composition of a painting on its sales price, e.g., \citet{stepanova2019impact} and \citet{ma2022colors}. The idea is that certain colors may trigger specific emotions, consequently influencing the price paid.\footnote{For example, \citet{bellizzi1992environmental} demonstrate the effects of various interior colors in department stores on consumer purchase intentions and buying behavior.} This paper, however, will approach the image analysis of artworks from a different perspective. We try to propose an objective measurement for the quantity of information that a painting delivers and see if it constitutes a significant factor in determining its sales price. 

Our main premise is that the valuation of an artwork is fundamentally influenced by the degree of consensus among professionals. So, any factor under consensus tend to be a potential pricing component, either positively or negatively. We know that a painting's content is the most important component of the artwork, while it is also the most difficult component for reaching consensus because people have quite subjective preferences. That is why the currrent research normally adopts the objective features of a painting into the hedonic model, since an objective feature can always obtain unambiguous consensus by all people. Such features include a painting's physical attributes (e.g., size and material), content attributes (e.g., signature, date and color compositions) and economic attributes (e.g., auction house and auction date). 

This paper focuses on the information extraction from the painting content. We wish to propose a new content variable that improves on the ususal ones. 
We can see that the signature and dated dummies provide rather coarse information on the painting content, while the color compositions are not quite robust as we show later. We will demonstrate that the information quantity of a painting can uniquely label the painting's content, and it proves to be an objectve, significant and robust pricing component.      

In order to measure the information quantity, we need to define ``information" in the first place. Information is a broad concept, any fact, idea, meaning and so on can all be called information. This paper will only focus on the narrow information concept defined by \citet{shannon1948entropy}, where the information is a means to resolve uncertainty. Therefore, a message that resolves greater uncertainty contains more information. \citet{shannon1948entropy} also suggested a measurement for the quantity of information, i.e., the entropy. 

Entropy is the fundamental concept in information theory. It is the kind of irreducible complexity below which a signal (message) cannot be further compressed. Mathematically, it is defined as the sum of the probability of each state of the uncertainty from a signal multiplied by the log probability of that same state. It is used to measure the quantity of information in general settings. In the economics and finance literature, entropy also finds its applications. For example, \citet{cabrales2013entropy,cabrales2017normalized} uses Shannon entropy to measure the informativeness of different information structures for investors when making investment decisions. \citet{sims2006rational} and \citet{miao2022multivariate} adopt the Shannon entropy to measure uncertainty and information capacity in the study of rational inattention.

Since an artwork transmits information, we can naturally use entropy to measure the amount of information transmitted. We will rely on an artwork's digital image because all the information is encoded within the image's pixels.\footnote{Note that the pixel is the basic unit of a digital image. Generally, more pixels in a given area of an image correspond to a higher image resolution.} The arrays of pixels constitute lines and shapes, and the combinations of pixel hues form colors. Therefore, the pixels not only carry information about their hue attributes but also their location coordinates in the image. This means that every pixel contains multidimensional information, and each dimension is indispensable to the whole image. For example, if we change some pixels' locations (or hues) in an image, we will change the picture altogether.

How can we measure this kind of multidimensional information given that traditional entropy is often used to quantify one-dimensional information? For this purpose, we introduce singular value decomposition (SVD) entropy.\footnote{The SVD entropy first finds the singular values of the pixel matrix and then calculates the entropy of these singular values. We will explain it in more detail in Section~\ref{sec:measure}.}  SVD is a common tool to extract important features from an image. It is often used to reduce the dimensionality of data. The SVD entropy is the kind of entropy defined over the one-dimensional singular values of the pixel matrix. We will calculate it based on the painting's color-removed image, which we define as the ``backbone information" of the painting. We then introduce this new measure into the traditional hedonic regression model and test its significance.

We obtained more than half a million observations for our empirical analysis. Our benchmark regressions show that the information quantity of an artwork positively affects the sales price at 1\% significance level in various specifications. We also run the robustness tests. We control for the topics of artworks in the regression, and the information quantity measurement remains significant. The subsample regression for each topic also provides similar results. Controlling the artwork styles does not change the result either. 

This positive influence of information quantity on the painting's sales price is intuitive. The literature generally finds that a signed or dated painting can fetch a price premium, ceteris paribus. Although coarse, these dummies still provide some kind of additional information, hence resolving some uncertainties. It is generally recognized that an investment asset with more uncertainties tends to have a lower price so that it can yield a risk premium. The same mechanism applies to the information quantity measurement. According to Shannon's definition, information resolves uncertainty, and more information resolves more uncertainties. Therefore, a painting that delivers a larger amount of information is associated with a lower level of uncertainty, conseqently booting its sales price.

We finally compare the SVD entropy to the other content variables. We show that the SVD entropy is more robust than the color compositions, which are sensitive to the testing samples. The SVD entropy is also superior to the signature and dated dummies in terms of the improvement of model fit (measured by adjusted R squared). The SVD entropy is easy to calculate as long as we have access to the artwork's digital image. Therefore, it is convenient to include this measurement in the usual hedonic regression for a better inference. From a practical point of view, our method may also provide a new perspective for the valuation of artworks.

The remainder of the paper is organized as follows. Section~\ref{sec:background} lays out the theoretical background. Section~\ref{sec:measure} describes the information quantity measurement in detail, and Section~\ref{sec:data} explains our data and methodology. In Section~\ref{sec:result}, we provide the empirical results of benchmark regression and robustness tests. Section~\ref{sec:application} compares the information quantity measurement to the other content variables and discusses its applicability. Section~\ref{sec:conc} concludes.

\section{Theoretical background and hypothesis} \label{sec:background}

This section will discuss the theoretical background and explain how it leads to our main hypothesis.  

\subsection{Information extraction from paintings}

Our research analyzes digital images of paintings. The digital image is composed of pixels. Each pixel is a basic information capsule that can be completely describled by the pixel's R (red), G (green) and B(blue) values, and the pixel's two-dimensional coordinates in the image plane.\footnote{RGB stands for the strength of the three additive primary colors (red, green and blue) in a pixel on the scale of 0 to 255. For example, RGB (0, 0, 0) means that the proportions of red, green and blue are all 0\%, so the pixel is black. RGB (255, 255, 255) indicates that the proportions of the three colors are 100\%, and the pixels are white.} Researchers try to extract useful information from these painting images in order to better understand the price-formation mechanism of paintings.

The literature in this strand mainly concentrates on the relationship between a painting's color composition and its sales price. For example, \citet{pownall2016pricing} average the R, G and B values of all pixels in a painting image and introduce them into the traditional hedonic model as the new explanatory variables. \citet{stepanova2019impact} further integrates this three-dimensional RGB information via a clustering algorithm, and identifies those principal color clusters that may affect a painting's sales price. \citet{ma2022colors} study the effect of colors on artwork pricing via the HSV (hue, saturation, value) color system. HSV describes the three different aspects of color, each of which is a unique aggregation of the underlying RGB values.\footnote{See Appendix~\ref{appx a} for detailed formulas.} Compared to RGB values, the HSV system appears more intuitive considering the human perception of colors.

The literature also raises two main concerns for us. First, previous research reveals certain links between colors and human emotions, e.g., \citet{boyatzis1994children} and \citet{hemphill1996note}. But how emotions can affect investment decisions seems more complex. There are both rational and behavioral schools in the study of investment decisions. To disentangle the effect of emotions during the art investment decision process poses great challenges. Further research is warranted for this color-emotion-investment transmission mechanism.

Second, even though HSV is more intuitive than RGB, it does not reduce the dimensionality of information. How this multidimensional information interacts with each other to influnce art pricing also calls for further studies.\footnote{Besides HSV, there exist other alternative color models like HSI (hue, saturation, intensity), HSL (hue, saturation, lightness), HSB (hue, saturation, brightness), etc. Whether the effect of colors on art pricing is sensitive to the choice of color models also needs further studies.} More importantly, a painting's colors are intrinsically attached to its structural elements, such as line, shape, form, space, etc. This implies that the \textit{positional} information of color element in the image plane shall also be considered, which, however, is often ignored in the previous research. A further complication also arises from the interaction of color elements with these structural elements when they contribute to the sales price aggregately.

\citet{ma2022colors} address some of these issues. For example, to separate the effect of colors from the structural elements of a painting, they adopt a sample that excludes those with figurative work, repeated patterns and less traditional shapes. To test the color-emotion-valuation mechanism, they run lab experiments in different locations, collecting the viewers' subjective valuations for the images with different color attributes. We think the experiment results might be interpreted with some caution, because the actual sales price is driven by the real buyer's decision (not the viewer's). Interestingly, \citet{blum2021price} finds that the tastes of collectors often go in the opposite direction to those of outsiders. 

From the above discussion, we can see that predicting a painting's sales price based on its color component requires certain restrictions. In this paper, we then ask whether other types of information can be extracted from a painting's image, which is more robust and applicable to more heterogenous painting contents. Drawing on the Shannon information theory, we suggest a concept called \textit{information quantity}, as a new type of information extraction for paintings. The following discussion tries to answer two essential questions: the first is why the information quantity of a painting can be linked to its sales price, and the second is what direction this link is.

\subsection{Consensus on information quantity}

The valuation of a painting, as well as other cultural goods, is often considered to lack objective criteria. In contrast, for those traditional assets like stocks and bonds, we can always benchmark against their dividends and interests. In spite of the existence of generally agreed guidelines, the actual assessment of cultural goods often involves subjective judgements. When an artwork receives conflicting views, its valuation must rely on some kind of consensus reached by specific groups through social interactions (\citet{lamont2009professors}, \citet{lewandowska2020striving}).

Then, what kind of attributes of artwork can share the agreement of all people? For a painting, we often point to those \textit{objective} features, e.g., the painting size and material. They have long been included in the hedonic regression model. So we propose that those features of unambiguous consensus tend to act as significant pricing factors for paintings. 

Of course, the most essential part of a painting is its content. But people find it much harder to reach a consensus when interpreting an artwork's content, because people are born with differential preferences (\citet{graham2010preference}, \citet{blum2021price}).  We suggest looking for those features in a painting's content that can be described objectively, hence obtaining a natural consensus. Some of such variables that we know include the painting's signature, date, color composition, etc. Not surprisingly, various researchers have already introduced them into the art pricing model. Note that the signature and date are dummy variables hence providing very coarse information on content. The color composition is unique to each painting, but it requires additional restrictions to be significant hence not robust (as we argued before and will show below). 

We propose the amount of information (information quantity) contained in the artwork image as a new content variable. It quantifies how much information a painting delivers to the audience. It uniquely labels each painting's content and can be measured objectively. So it also belongs to the features of unambiguous consensus, similar to size, material, signature, etc. Naturally, it could also be a potentially significant pricing factor.  

In this paper, we adopt a narrow definition for information proposed by Claude Shannon, i.e., a message sent by a sender to a receiver. He suggested a quantity measurement for such information, defined as entropy. Shannon asserted that information is used to resolve uncertainty. Therefore, the amount of information a signal transmits can be quantified by the degree of uncertainty that the signal resolves. A higher degree of uncertainty from a signal implies a greater chance of `surprise' by the receiver, hence resolving it delivers more information. More rigorously, Shannon's entropy is defined as the sum of the probability of each state of the uncertainty multiplied by the log probability of that same state, that is, 
$$ E = -\sum_{i=1}^{n}p_i \log_2 p_i, $$
where $p_i$ is the probability of state $i$ among all the $n$ possible states for a signal.

Shannon's entropy formula is independent of the meaning of the message. In fact, \citet{shannon1948entropy} suggested, ``information in the form of a message often contains meaning, but that meaning is not a necessary condition for defining information. It is possible to have information without meaning, whatever that means.''  When entropy avoids subjective interpretation completely, it becomes a consensus automatically. We can think of the content entropy for a painting similarly to the number of bytes for a file, which is an objective measurement too.  

We hypothesize that the Shannon information quantity (measured by entropy) would contribute to the sales price of a painting positively. Although a painting is not the same as a telegram, an analogy from the traditional telegram message may provide us some intuition. A telegram message is usually charged by the number of words used (objective) rather than the meaning or importance of the message (subjective). So statistically, we would expect a larger information quantity corresponds to a higher sales price of the painting, ceteris paribus.

The underlying mechanism is similar to how the signature or the dated dummy affects the painting's price. The literature generally finds that a painting with signature or date information on its image can obtain a higher auction price than those without, ceteris paribus. The reason is that this additional information resolves some kind of uncertainties about the painting, hence the buyer is willing to accept a higher bid for bearing a lower risk. This is a common principle in efficient financial market, which applies to the art market as well. Therefore, a painting with a higher Shannon information quantity transmits more information, hence resolving more uncertainties in one way or another. Consequently, a higher price is expected, ceteris paribus.

\section{Measurement of information quantity in an artwork} \label{sec:measure}

A digital image with a resolution of $m\times n$ has $m$ pixels in the horizontal and $n$ pixels in the vertical direction of the image. So the total number of pixels is $m\, n$. Then, the information contained in the image must be some kind of aggregation of the information from all $m\, n$ pixels.\footnote{Note that a higher resolution does not necessarily imply a larger information quantity. Shannon information quantity emphasizes the variation of the pixels rather than the absolute number of pixels. } 

Our approach is to first strip off the colors from each pixel. Considering colors will increase the dimensionality of measurement tremendously as we discussed before. When colors are removed, the whole picture becomes black-and-white, and each pixel can be uniquely described by its coordinates  $(i, j)$ in the two-dimensional image space and its grayscale value, denoted as $grayscale_{ij}$. This information of all pixels will eventually be integrated into a one-dimensional measurement of the information quantity for the whole image.

Note that the grayscale of a pixel is defined as the weighted sum of its primitive RGB values.\footnote{The NTSC (National Television Standards Committee) formula for the grayscale is $0.3 R + 0.59 G + 0.11 B$.} For a black-and-white image, the R,G and B of a given pixel are equivalent, which all equal the pixel's grayscale value. Moreover, different color models, regardless of HSV, HSL or HSI, become equivalent as well. Therefore, we can call the information in an image after its colors are removed the \textit{backbone information} of the image. We will work on this backbone information to obtain our measurement of information quantity. 

Consider a digital image with $m \times n$ pixels. Its backbone information can be completely described by a $m \times n$ matrix with entry $grayscale_{ij}$, where $i \in \{1,2,...m\}$ and $j \in \{1,2,...n\}$. We denote this matrix as $M$. Then how can the information carried by $M$ be measured in an integrated way? We adopt a one-dimensional measurement called SVD entropy to account for the information quantity contained by
matrix $M$.

The singular value decomposition (SVD) of $M$ is given by $$ M = U\Sigma V^T,$$
where $U$ and $V$ are orthonormal matrices of size $m\times m$ and $n\times n$, respectively, and $\Sigma$ is an $m\times n$ diagonal matrix. The diagonal entries $\sigma_i = \Sigma_{ii}$ for $i=1,\ldots,p=\min(m,n)$ are called the singular values of $M$. The singular values are nonnegative and in decreasing order,
$$\sigma_1 \ge \sigma_2 \ge \cdots \ge \sigma_p \ge 0.$$
The number of nonzero singular values corresponds to the rank of the matrix $M$.\footnote{Note that the special case of SVD applied to the covariance matrix is well known as principal component analysis (PCA). Since the covariance matrix is symmetric (and positive semidefinite), $U=V$ ($m=n$), and the singular values are also the eigenvalues of the covariance matrix.} 

With SVD, $M$ can always be decomposed as a weighted sum of separable matrices,
\begin{equation} \label{eq:sep}
	M = \sum_{i=1}^p \sigma_i \, U_i V_i^T,
\end{equation}
where $U_i$ and $V_i$ are the $i$-th column vectors of $U$ and $V$, respectively. Here, $U_i V_i^T$ is a separable $m\times n$ matrix with intensity $\sigma_i$. SVD is a common tool to reduce the dimensionality of the data. The truncated SVD using the first $r$ terms,
$$ \tilde{M}_r = \sum_{i=1}^r \sigma_i \, U_i V_i^T \quad\text{for}\quad r<p,
$$
is the best approximation of the original matrix $M$ among the matrices with rank $r$. The truncated SVD is often used to compress images into reduced storage.

In Equation~\eqref{eq:sep}, the singular value $\sigma_i$ is understood as the `probability` of a `state' $U_i V_i^T$ of the matrix $M$. We can naturally define the entropy out of the singular values $\sigma_i$. The SVD entropy of $M$ is given by the entropy of the normalized singular values:
$$ E(M) = - \sum_{i=1}^p \bar{\sigma}_i \log_2 \, \bar{\sigma}_i, \quad \text{where} \quad
\bar{\sigma}_i = \frac{\sigma_i}{\sum_{k=1}^p \sigma_k}.
$$
The mathematical property of the SVD entropy is similar to that of the traditional Shannon entropy. The possible minimum value of the SVD entropy is zero, $E(M)=0$, when there is only one nonzero singular value.\footnote{We assume that $0 \cdot \log 0 = \lim_{x\rightarrow 0} x \log_2 x = 0$} For example, a monotone image (e.g., all pixels with the same grayness) has zero SVD entropy. The possible maximum value is $E(M)=\log p$ when the singular values are all the same as a nonzero value. The maximum value is achieved when a painting is represented by any orthonormal matrix.

\begin{figure}[ht!]
	\caption{`Composition with Red, Blue and Yellow' by Piet Mondrian (left), `Number 5' by Jackson Pollock (right) in grayscale, and the distributions of their normalized singular values (bottom).
		%The SVD entropy values of the two paintings are 3.91 and 7.96, respectively. For SVD, the paintings are rescaled to $400\times 400$ in grayscale. 
		\label{f:examples}}
	\centering
	\includegraphics[width=0.4\textwidth]{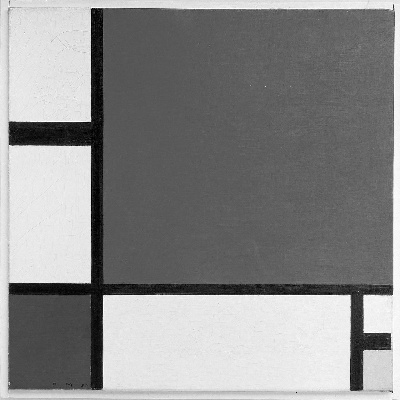}\hspace{2em}
	\includegraphics[width=0.4\textwidth]{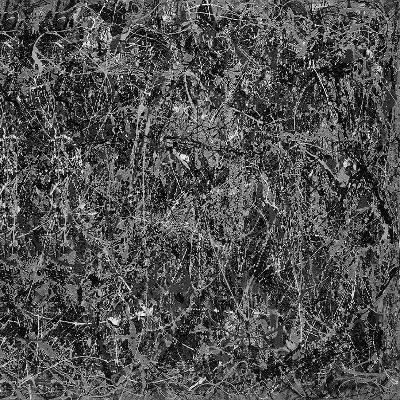}\\ \vspace{2ex}
	\includegraphics[width=0.4\textwidth]{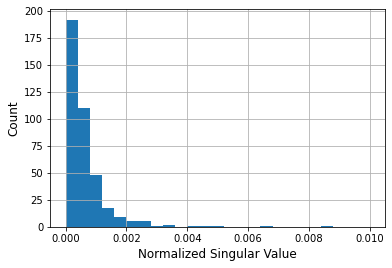}\hspace{2em}
	\includegraphics[width=0.4\textwidth]{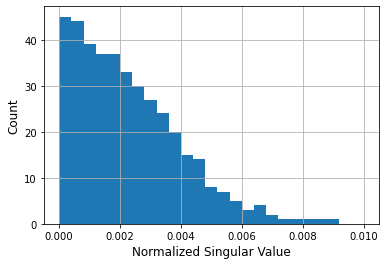}\\
	%\begin{figurenotes}
	\vspace{2ex}
	\footnotesize
		\qquad Note: Figure~\ref{f:examples} plots the distributions of the normalized singular values (bottom) of `Composition with Red, Blue and Yellow' by Piet Mondrian and `Number 5' by Jackson Pollock. The grayscale SVD entropy  values of the two paintings are 3.91 and 7.96. For SVD, the paintings are rescaled to $400\times 400$.
	%\end{figurenotes}
\end{figure}

Figure~\ref{f:examples} illustrates the singular value distribution and SVD entropy with two paintings: \textit{Composition with Red, Blue and Yellow} by Piet Mondrian (left) and \textit{Number 5} by Jackson Pollock (right). They are both transformed from the original color images into the gray ones. The two paintings show a stark contrast in complexity. \textit{Composition with Red, Blue and Yellow} has an SVD entropy 3.91, as its singular values are concentrated near zero. However, \textit{Number 5}  has an SVD entropy of 7.96, as its singular values are widely spread out. This SVD entropy value is close to the maximum possible value, $\log_2 400 = 8.64$. 

SVD entropy has several properties suitable for computing the information quantity based on digital images. For example, singular values are invariant under both rotation (by $90^\circ$, $180^\circ$, and $270^\circ$) and reflection (on the $x$ and $y$ axes). Additionally, SVD entropy is invariant under the linear scaling of the image matrix (e.g., brightness adjustment). Because the singular values of $\lambda M$ are $\lambda\sigma_i$ ($\lambda>0$), the SVD entropy of $\lambda M$ is the same as that of $M$.

SVD entropy is widely applied in feature selection research in computer graphics, e.g., \citet{banerjee2014feature} and \citet{buisine2021stopping}. SVD effectively extracts the important features from a painting, so the entropy based on these important features, i.e., the SVD entropy, will be an appropriate measurement for the painting's information quantity. 

\section{Data and methodology} \label{sec:data}

Next, we will describe our data and empirical methodology, which we adopt to test the effect of SVD entropy on a painting's sales price.   

\subsection{Data}

We obtain our data from \href{www.findartinfo.com}{FindArtInfo}, an art database website that provides auction records of artwork worldwide from 2000 to 2015. The database has a large number of records with auction prices, painting attributes, and digital images of auctioned items, making it ideal for our research. A lot of recent research, such as \citet{filipiak2016towards}, \citet{ayub2017art}, and \citet{powell2019developing}, has adopted this database. In particular, we collect auction records from the website using the web crawler algorithm modified from that used in \citet{ayub2017art}.\footnote{\url{https://github.com/orbancedric/CS229-Final-Project}}

The database contains more than one million auction records of artwork in various forms from approximately 300,000 artists. To focus on the artworks with liquidity, we restrict our sample to the auction items that belong to the top 1\% artists, who are ranked by the number of auction records (with images) in the database. We then obtain 3,004 artists, whose minimum number of observations is 99. Among all forms of artwork, we only select those related to paintings via their medium keywords as the search criteria.\footnote{We eliminate those records with the following keywords in their medium labels: bronze, iron, sculpture, terracotta, assemblage, steel, marble, aluminum, brass, ceramic, porcelain, resin, plaster, metal, plastic and stone.} Finally, we obtain 529,654 observations with digital images. Although they are only from the top 1\% artists, the sample accounts for approximately 38\% of the total records in the database.  Table~\ref{tab:p_dist} reports the distribution of the auction prices. We can see that most of the auction prices (75.44\%) fall into the range between 100 and 100,000 US dollars. 

\begin{table}[htp!] \small\centering
	\caption{Sample price distribution \label{tab:p_dist}}
	\begin{tabular}{crr}
		\hline
		\hline
		Price          & $N\quad$ & Percentage  \\ \hline
		{[}0, 100)         & 18,480 & 3.49\% \\ 
		{[}100, 1,000)      & 171,639 & 32.41\% \\ 
		{[}1,000, 10,000)     & 227,886 &  43.03\% \\ 
		{[}10,000, 100,000)    & 90,556 &  17.10\% \\ 
		{[}100,000, 1,000,000)  & 18,372 &   3.47\%\\ 
		{[}1,000,000, 10,000,000) & 2,505  &  0.47\%\\ 
		{[}10,000,000, $\infty$)    & 216  &   0.04\%\\ 
		\hline
		Total  & 529,654 &100.00\%\\
		\hline\hline %\vspace{0ex}	
	&&\\
%\begin{tablenotes}
		\multicolumn{3}{l}{\footnotesize{\emph{Note: }The auction price is converted into US dollars with the exchange}}\\
		\multicolumn{3}{l}{\footnotesize{rate in the corresponding auction year.}}
%\end{tablenotes}
\end{tabular}
\end{table}

When retrieving the digital images, we resize them to uniform $400 \times 400$ pixels using the bicubic interpolation method rather than using the images in their original resolutions.\footnote{See \citet{keys1981cubic} for the introduction of bicubic interpolation.} This is because we want to compare the SVD entropies when the physical dimensions of the painting are controlled. Note that the width and height of paintings are to be included as control variables in all of our regression specifications as well.

\subsection{Variables and the hedonic regression}

From each normalized image, the grayscale SVD entropy, denoted as $E_g$, is calculated according to the description in Section~\ref{sec:measure}. For the whole sample, the mean of $E_g$ is 5.18, with a standard deviation of 1.08. The maximum value of $E_g$ is 8.04, and the minimum is 1.07e-12. Table~\ref{tab:sta of var} provides the summary statistics of all the variables for our empirical study. 

Besides the dependent variable auction price, and the key explanatory variable SVD entropy, Table~\ref{tab:sta of var} lists all the other control variables, such as painting size, artist's signature and date, medium, artist, auction house and city, and auction year and month. We can see that the average auction price is 37,812 USD and that the price dispersion appears to be quite large. Note that we have already converted auction prices to inflation-adjusted USD.\footnote{We use 2000 as the base year for the inflation adjustment.} The average height and width of the paintings are both approximately 19 inches. 72\% and 27\% of the paintings in the sample are signed and dated, respectively. 

For each dummy variable in Table~\ref{tab:sta of var}, we show the statistics of one representative value while omitting the rest for brevity. For example, the artist dummy differentiates all 3,004 artists. We only supply the statistics for Pablo Picasso, whose observations are the most in the sample, reaching 6,589. When we construct the dummies for the variables like  medium, auction house, and auction city, we condense those minor categories for each variable because the classification provided by the database is too fine. For these variables, we first rank the categories by their frequencies in the sample. We then take the top 50 mediums, the top 20 auction houses, and the top 20 auction cities. For the rest of the minor classes, we group them into the `\textit{other}' category for each variable. The $other$ categories in medium, auction houses, and auction cities account for 30.60\%, 51.21\%, and 36.93\% of the sample respectively. Finally, the year and month dummies are also introduced to control for the time effect of auction.\footnote{Note that the year dummy does not include year 2001 to 2003, because our sample with top 1\% artists has no auction records during this period.} 

We apply the following hedonic regression model to test the effect of the information quantity in a painting's image on its sales price:
\begin{equation} \label{eq:reg}
	\log \, p_{i} = \alpha E_i + \sum_{k=1}^K\beta_{k}X_{ki} + 	\sum_{t=1}^T\gamma_{t}D_{it}+\epsilon_{i}, 
\end{equation} 
where $E_i$ is the set of information quantity measurements (i.e., the SVD entropy and its square term) of painting $i$, $X_{ki}$ is the set of time-invariant characteristics of painting $i$ (e.g., the height, width, medium dummy, signature dummy), and $D_{it}$ is the set of time varying idiosyncratic attributes (e.g., year dummy and month dummy) of painting $i$. The logarithm of the inflation-adjusted price $p_{i}$ of painting $i$ sold at the auction  is used as the dependent variable in the regression.
\begin{table}[htp!] \small\centering
\caption{Descriptive statistics of regression variables 
\label{tab:sta of var}}
\begin{tabular} {@{}lcrrrr@{}} \hline\hline
	Variable  &     $N$ &    Mean\ \ &     Sd\quad\ &    0\ &  1\ \\  \hline
	Price (USD) & 529,654 &37,812.05 & 554,851.84 & & \\
	SVD Entropy ($E_g$) & 529,654 &5.19 & 1.08 & & \\
	[1em]		
	Height (inch) & 529,654 & 19.49 & 28.85 & & \\
	Width (inch) & 529,654 & 19.90 & 35.21 & & \\
	Signature dummy &529,654 &    0.72 &    0.45 &   147,857 &   381,797 \\
	Dated dummy &529,654 &    0.27 &    0.45 &    384,351 &  145,303 \\		
	[1em]
 	Artist dummy & &&&&\\ \cline{1-1}
	Pablo Ruiz Picasso	&	529,654	&	0.01 	&	0.01 	&	523,065	&	6,589	\\
	Other artists & omitted & omitted & omitted & omitted & omitted \\
	[1em]
	Medium dummy & &&&&\\ \cline{1-1}
	Oil on canvas	&	529,654	&	0.16 	&	0.13 	&	444,583	&	85,072	\\
	Other media & omitted & omitted & omitted & omitted & omitted \\	[1em]
	Auction house dummy & &&&&\\ \cline{1-1}
	Christie's	&	529,654	&	0.13 	&	0.11 	&	461,501	&	68,153	\\
	Other houses & omitted & omitted & omitted & omitted & omitted \\	[1em]
	Auction city dummy & &&&&\\ \cline{1-1}
	New York	&	529,654	&	0.13 	&	0.11 	&	462,101	&	67,553	\\
	Other cities & omitted & omitted & omitted & omitted & omitted \\	[1em]
	Year dummy & &&&&\\ \cline{1-1}
	2000	&	529,654	&	0.00 	&	0.00 	&	529,647	&	7	\\
	Other years & omitted & omitted & omitted & omitted & omitted \\	[1em]
	Month dummy & &&&&\\ \cline{1-1}
	January	&	529,654	&	0.03 	&	0.03 	&	513,078	&	16,576	\\
	Other months & omitted & omitted & omitted & omitted & omitted \\
	\hline\hline
%	\multicolumn{6}{@{}l}{\footnotesize{\emph{note:} }}
&&&&&\\
%\begin{tablenotes}
	\multicolumn{6}{l}{\footnotesize{\emph{Note: }Table~\ref{tab:sta of var} presents the descriptive statistics of representative variables in our model. }}\\
	\multicolumn{6}{l}{\footnotesize{For brevity, we omit most dummies in this table. }}
%\end{tablenotes}
\end{tabular}
\end{table}

\section{Empirical results} \label{sec:result}

We first provide the empirical results for the benchmark regression according to Equation~\eqref{eq:reg}. Then we run various robustness tests by considering the topics and styles of the paintings. 

\subsection{Benchmark regression}

Table~\ref{tab:benchmark} presents the results for the benchmark regressions according to Equation~\eqref{eq:reg} in various specifications. The SVD entropy $E_g$ positively affects the sales price at 1\% significance level for all specifications. The coefficient of the quadratic term of SVD entropy $E_g^2$ is also significantly positive. The cross products of the SVD entropy and the painting's height and width are both significantly negative at very small magnitudes, implying a slightly weaker influence of $E_g$ on the price for a bigger painting. All the results demonstrate that the information quantity, measured by the SVD entropy, can indeed be a significant predictor of the painting's sales price. 

The coefficients and significance of other variables are consistent with the previous literature. For example, both the height and the width of a painting significantly enhance its sales price but with a diminishing effect. The existence of signature and date in a painting's image also significantly increases the painting's value. Note that we omit the results for most dummy variables in Table~\ref{tab:benchmark}. We put more detailed outputs in Appendix~\ref{appx b}.

\begin{table}[htp] \footnotesize\centering
\caption{The benchmark hedonic regressions. \label{tab:benchmark}}
\begin{tabular}{l*{5}{c}}
	\hline\hline
	&\multicolumn{1}{c}{(1)}&\multicolumn{1}{c}{(2)}&\multicolumn{1}{c}{(3)}&\multicolumn{1}{c}{(4)}&\multicolumn{1}{c}{(5)}\\
	
	\hline
	$E_g$          &                     &       0.102\sym{***}&       0.021\sym{**} &       0.110\sym{***}&       0.107\sym{***}\\
	&                     &     (0.002)         &     (0.010)         &     (0.002)         &     (0.002)         \\
	[1em]
	$E_g^2$         &                     &                     &       0.008\sym{***}&                     &                     \\
	&                     &                     &     (0.001)         &                     &                     \\
	[1em]
	Height      &       0.006\sym{***}&       0.006\sym{***}&       0.007\sym{***}&       0.009\sym{***}&       0.006\sym{***}\\
	&     (0.000)         &     (0.000)         &     (0.000)         &     (0.000)         &     (0.000)         \\
	[1em]
	Height$^2$      &      -0.000\sym{***}&      -0.000\sym{***}&      -0.000\sym{***}&      -0.000\sym{***}&      -0.000\sym{***}\\
	&     (0.000)         &     (0.000)         &     (0.000)         &     (0.000)         &     (0.000)         \\
	[1em]
	$E_g$*\hspace{0.05cm}Height   &                     &                     &                     &      -0.000\sym{***}&                     \\
	&                     &                     &                     &     (0.000)         &                     \\
	[1em]
	Width       &       0.006\sym{***}&       0.006\sym{***}&       0.006\sym{***}&       0.006\sym{***}&       0.007\sym{***}\\
	&     (0.000)         &     (0.000)         &     (0.000)         &     (0.000)         &     (0.000)         \\
	[1em]
	Width$^2 $      &      -0.000\sym{***}&      -0.000\sym{***}&      -0.000\sym{***}&      -0.000\sym{***}&      -0.000\sym{***}\\
	&     (0.000)         &     (0.000)         &     (0.000)         &     (0.000)         &     (0.000)         \\
	[1em]
	$E_g$*\hspace{0.05cm}Width    &                     &                     &                     &                     &      -0.000\sym{***}\\
	&                     &                     &                     &                     &     (0.000)         \\
	[1em]
	Signature dummy  &       0.140\sym{***}&       0.139\sym{***}&       0.139\sym{***}&       0.139\sym{***}&       0.139\sym{***}\\
	&     (0.004)         &     (0.004)         &     (0.004)         &     (0.004)         &     (0.004)         \\
	[1em]
	Dated dummy &       0.181\sym{***}&       0.176\sym{***}&       0.176\sym{***}&       0.176\sym{***}&       0.176\sym{***}\\
	&     (0.004)         &     (0.004)         &     (0.004)         &     (0.004)         &     (0.004)         \\
	[1em]
	Artist dummy   &  control   & control     &   control   &   control     &  control       \\
	[1em]
	Medium dummy&   control   & control     &   control   &   control   &     control         \\
	[1em]
	Auction house dummy&   control   &  control    &   control   &  control     &  control        \\
	[1em]
	Auction city dummy&  control   & control    &   control    &   control     &  control         \\
	[1em]
	Year dummy&  control     &  control    &  control    &     control       &  control         \\
	[1em]
	Month dummy&  control    &   control    &  control     &     control     &  control        \\
	[1em]
	\hline
	\(N\)    &   529,654     &   529,654     &   529,654     &   529,654     &   529,654         \\
	Adj. \(R^{2}\)&       0.657         &       0.659         &       0.659         &       0.659         &       0.659         \\
	\hline\hline
&&&&&\\
%\begin{tablenotes}\\
	\multicolumn{6}{l}{\footnotesize{\emph{Note: }1. Table~\ref{tab:benchmark} presents the results of the hedonic regressions of artworks' log prices on }}\\ 
	\multicolumn{6}{l}{\footnotesize{the information quantity measurements and other control variables. More details are shown }}\\
	\multicolumn{6}{l}{\footnotesize{in Appendix~\ref{appx b}.}}\\
	\multicolumn{6}{l}{\footnotesize{2. The number of observations ($N$) and the adjusted R-squared (Adj. $R^2$) are presented at the}}\\
	\multicolumn{6}{l}{\footnotesize{bottom of the table.}}\\
	\multicolumn{6}{l}{\footnotesize{3. Standard errors are in parentheses.}}\\
	\multicolumn{6}{l}{\footnotesize{4. ***, **, and * denote significance at the 1\%, 5\%, and 10\% levels, respectively.}}
%\end{tablenotes}
\end{tabular}
\end{table}

\subsection{Robustness with topic} \label{ss:topic}
In the art pricing literature, the topic and style of a painting have been used to explain content heterogeneity. We hold that the SVD entropy is a measure independent of topic and style and that the explanatory power of SVD entropy remains strong even when they are controlled.

We first check if our results are robust against the topics, i.e., the subject matters. \citet{renneboog2012buying} classify the topics of paintings into ten large categories: abstract, animals, landscape, nude, people, portrait, religion, self-portrait, still life, and urban. Their classification strategy is implemented by matching the first few words in the title of a given painting with the set of keywords defined for each topic.\footnote{See Appendix~\ref{appx c} for the detailed keywords table.} Following the same methodology, we successfully obtain those records that can be labeled by topic, which account for approximately 23\% of our original sample. We focus on this sample of 123,221 observations and construct topic dummies based on the above classification criteria.

The ten topic dummies correspond to ten subsamples of distinct topics. Table~\ref{tab:desstat} provides the descriptive statistics of the SVD entropy for each of them. We can observe that the standard deviations are typically larger than the differences in the means. It indicates that the SVD entropy adds new information different from that provided by the topic. Among all the topics, we can see the SVD entropy of \textit{nude} has the smallest mean but the largest standard deviation. In contrast, the entropy of \textit{still life} has the largest mean while the smallest standard deviation. So we may infer that the paintings that depict \textit{still life} have much smaller content variations than those depicting \textit{nude} in terms of how much information is delivered.

\begin{table}[htp!] \small\centering
	\caption{Descriptive statistics of the SVD entropies for different topics.
		\label{tab:desstat}}
	\begin{tabular}{lrcccc}
		\hline
		\hline
		&  & \multicolumn{4}{c}{$E_G$}\\ \cline{3-6} 
		Topic & \multicolumn{1}{c}{$N$}     & Mean     & Sd       & Min      & Max      \\		\hline
		Abstract      & 13,581  & 5.410 & 1.081 & 0.752 & 7.897  \\
		Animals    & 8,002   & 5.172 & 0.984 & 0.670 & 7.723 \\
		Landscape     & 21,030  & 5.196 & 0.973 & 0.790 & 7.960  \\
		Nude         & 5,631   & 4.661 & 1.213 & 0.725 & 7.514 \\
		People          & 30,218  & 5.079 & 1.087 & 0.231 & 7.856  \\
		Portrait          & 9,042   & 4.951 & 1.040 & 0.780 & 7.437  \\
		Religion       & 4,447   & 5.395 & 1.109 & 0.411 & 7.737    \\
		Self-portrait   & 1,865   & 5.103 & 1.034 & 0.582 & 7.673  \\
		Still life      & 6,896   & 5.536 & 0.866 & 1.013 & 7.559  \\
		Urban    & 22,509  & 5.324 & 0.956 & 0.355 & 7.786  \\
		\hline
		Total         & 123,221 & 5.195 & 1.047 & 0.230 & 7.960  \\
		\hline\hline
	&&&&&\\
	%\begin{tablenotes}
	
		\multicolumn{6}{l}{\footnotesize{\emph{Note: }Table~\ref{tab:desstat} presents the descriptive statistics of the SVD entropies}}\\
		\multicolumn{6}{l}{of grayscale values for different topic categories. }
	%\end{tablenotes}
\end{tabular}
\end{table}

We introduce the topic dummy $Topic_{i}$ as an additional control variable, and run regressions according to Equation~\eqref{eq:regtopic}.

\begin{equation} \label{eq:regtopic}
\log \, p_{i} = \alpha E_i + \sum_{k=1}^K\beta_{k}X_{ki} + 	\sum_{t=1}^T\gamma_{t}D_{it}+\delta\, Topic_{i}+\epsilon_{i}, 
\end{equation} 
Table~\ref{tab:topic1} presents the regression results. They are almost the same as the baseline regressions in Table~\ref{tab:benchmark}. For example, in Specification (5), the painting size also slightly negatively moderates the effect of the information quantity $E_g$.

\begin{table}[htbp] \scriptsize\centering
	\caption{The hedonic regressions with the topic dummy. }
	\label{tab:topic1}
	\begin{tabular}{l*{5}{c}}
		\hline\hline
		&\multicolumn{1}{c}{(1)}&\multicolumn{1}{c}{(2)}&\multicolumn{1}{c}{(3)}&\multicolumn{1}{c}{(4)}&\multicolumn{1}{c}{(5)}\\
		
		\hline
		$E_g$          &                     &       0.128\sym{***}&       0.128\sym{***}&       0.149\sym{***}&       0.149\sym{***}\\
		&                     &     (0.004)         &     (0.022)         &     (0.022)         &     (0.022)         \\
		[1em]
		$E_g^2$          &                     &                     &       0.000         &      -0.001         &      -0.001         \\
		&                     &                     &     (0.002)         &     (0.002)         &     (0.002)         \\
		[1em]
		Height      &       0.011\sym{***}&       0.011\sym{***}&       0.011\sym{***}&                     &                     \\
		&     (0.000)         &     (0.000)         &     (0.000)         &                     &                     \\
		[1em]
		Height$^2$     &      -0.000\sym{***}&      -0.000\sym{***}&      -0.000\sym{***}&                     &                     \\
		&     (0.000)         &     (0.000)         &     (0.000)         &                     &                     \\
		[1em]
		Width       &       0.008\sym{***}&       0.008\sym{***}&       0.008\sym{***}&                     &                     \\
		&     (0.000)         &     (0.000)         &     (0.000)         &                     &                     \\
		[1em]
		Width$^2$      &      -0.000\sym{***}&      -0.000\sym{***}&      -0.000\sym{***}&                     &                     \\
		&     (0.000)         &     (0.000)         &     (0.000)         &                     &                     \\
		[1em]
		Height\hspace{0.03cm}*\hspace{0.03cm}Width        &                     &                     &                     &       0.000\sym{***}&       0.000\sym{*}  \\
		&                     &                     &                     &     (0.000)         &     (0.000)         \\
		[1em]
		(Height\hspace{0.03cm}*\hspace{0.03cm}Width)$^2$       &                     &                     &                     &      -0.000\sym{**} &        \\
		&                     &                     &                     &     (0.000)         &              \\
		[1em]
		$E_g$*\hspace{0.03cm}Height\hspace{0.03cm}*\hspace{0.03cm}Width     &                     &                     &                     &                     &       -0.000\sym{*}         \\
		&                     &                     &                     &                     &     (0.000)         \\
		[1em]
		Signature dummy  &       0.171\sym{***}&       0.171\sym{***}&       0.171\sym{***}&       0.171\sym{***}&       0.171\sym{***}\\
		&     (0.008)         &     (0.008)         &     (0.008)         &     (0.008)         &     (0.008)         \\
		[1em]
		Dated dummy &       0.188\sym{***}&       0.182\sym{***}&       0.182\sym{***}&       0.200\sym{***}&       0.200\sym{***}\\
		&     (0.009)         &     (0.009)         &     (0.009)         &     (0.009)         &     (0.009)         \\
		[1em]
		\textit{\textbf{Topic dummy}}&  \textit{\textbf{control}}    &   \textit{\textbf{control}}    &  \textit{\textbf{control}}     &     \textit{\textbf{control}}     &  \textit{\textbf{control}}       \\
		[1em]
		Artist dummy   &  control   & control     &   control   &   control     &  control      \\
		[1em]
		Medium dummy&   control   & control     &   control   &   control   &     control        \\
		[1em]
		Auction house dummy&   control   &  control    &   control   &  control     &  control        \\
		[1em]
		Auction city dummy&  control   & control    &   control    &   control     &  control         \\
		[1em]
		Year dummy&  control     &  control    &  control    &     control       &  control         \\
		[1em]
		Month dummy&  control    &   control    &  control     &     control     &  control        \\
		[1em]
		
		\hline
		\(N\)       &      123,221         &      123,221         &      123,221         &      123,221         &      123,221         \\
		Adj. \(R^{2}\)&       0.670         &       0.673         &       0.673         &       0.660         &       0.660         \\
		\hline\hline
		&&&&&\\
	
	%\begin{tablenotes}\\
	
	\multicolumn{6}{l}{\footnotesize{\emph{Note: }1. Table~\ref{tab:topic1} presents the results of the hedonic regressions of artworks'}}\\
	\multicolumn{6}{l}{log prices on the information quantity measurements controlling for the topic and other}\\
	\multicolumn{6}{l}{variables. The  classification of topic dummy is explained in Appendix~\ref{appx c}.}\\
	\multicolumn{6}{l}{	2. The number of observations ($N$) and the adjusted R-squared (Adj.~$R^2$) are presented}\\
	\multicolumn{6}{l}{ at the bottom of the table.}\\
	\multicolumn{6}{l}{	3. Standard errors are in parentheses.}\\
	\multicolumn{6}{l}{	4. ***, **, and * denote significance at the 1\%, 5\%, and 10\% levels, respectively.}
	%\end{tablenotes}
\end{tabular}
\end{table}

Furthermore, we run the subsample regression for each topic category.  Table~\ref{tab:topic2} summarizes the coefficients of $E_g$, their significance levels, as well as the adjusted R squared from all the subsample regressions according to Specification (2) in Table~\ref{tab:benchmark}. For every topic, the SVD entropy positively affects the sales price very significantly.
When the magnitudes of the coefficients are compared, we observe that the SVD entropy has the largest impact on \textit{still life} while the smallest impact on \textit{nude}. This result is consistent with the statistical properties of SVD entropies from these two topic categories reported in Table~\ref{tab:desstat}.

\begin{table}[htp] \small\centering
	\caption{The subsample regressions by topics. \label{tab:topic2}}
	\begin{tabular}{llll|lllc}
		\hline\hline
		Topic && $N\quad$ &&& $\;\;E_g$ &&  Adj. $R^2$ \\
		\hline
		Abstract   && 13,581 &&& 0.117\sym{***}            && 0.740      \\
		Animals    && 8,002 &&& 0.117\sym{***}             && 0.691       \\
		Landscape   && 21,030 &&& 0.098\sym{***}          & & 0.739        \\
		Nude      && 5,631  &&& 0.063\sym{**}           &  & 0.730       \\
		People      && 30,218 &&& 0.147\sym{***}        &  & 0.680       \\
		Portrait   && 9,042 &&& 0.174\sym{***}          & & 0.653        \\
		Religion   && 4,447  &&& 0.105\sym{***}          & & 0.609       \\
		Self-portrait  && 1,865 &&& 0.158\sym{***}       &   & 0.687      \\
		Still life    && 6,896  &&& 0.190\sym{***}       &  & 0.748      \\
		Urban      && 22,509  &&& 0.094\sym{***}        & & 0.686     \\
		\hline\hline %\vspace{0ex}
	&&&&&&&\\
	%\begin{tablenotes}\\ 
	\multicolumn{8}{l}{\footnotesize{\emph{Note: }1. Table~\ref{tab:topic2} shows the regression results of Specification (2) in  Table~\ref{tab:benchmark} }}\\
	\multicolumn{8}{l}{using subsamples of different topics.  The classification of the topic}\\
	\multicolumn{8}{l}{dummy is explained in Appendix~\ref{appx c}.}\\
	\multicolumn{8}{l}{2. ***, **, and * denote significance at the 1\%, 5\%, and 10\% levels}\\
	\multicolumn{8}{l}{respectively.}
	%\end{tablenotes}
\end{tabular}
\end{table}

\subsection{Robustness with style}
\label{ss:style}

In addition to topic, style or movement is another content description for a painting. For example, \citet{renneboog2012buying} also classify paintings into thirteen styles, such as renaissance, baroque, and pop.\footnote{The thirteen styles in \citet{renneboog2012buying} include: medieval \& renaissance; baroque; rococo; neoclassicism; romanticism; realism; impressionism \& symbolism; fauvism \& expressionism; cubism, futurism \& constructivism; dada \& surrealism; abstract expressionism; pop; and minimalism \& contemporary.} As our main database \href{www.findartinfo.com}{FindArtInfo} does not provide the style label for an artwork item, we try to supplement this information from \href{www.wikiart.org}{WikiArt}, which has a large number of paintings with style labels but no prices. By matching the artist name and the painting title from these two databases, we obtain 21,027 observations whose auction prices and styles are both available. Note that this sample is much smaller than the previous sample with topic labels.

As \href{www.wikiart.org}{WikiArt} adopts a much finer style classification than \citet{renneboog2012buying},  our sample finally obtains more than eighty distinct styles.\footnote{Please refer to \href{www.wikiart.org}{WikiArt} for the detailed style classifications.} For brevity, Table~\ref{tab:style1} only presents the descriptive statistics of the SVD entropies for the top 10 styles ranked by the number of observations in the sample. The paintings of the top 10 styles account for 68.6\% of this sample.  
\begin{table}[htp!] \small \centering
	\caption{Descriptive statistics of the SVD entropies of Top 10 styles. \label{tab:style1}}
	\begin{tabular}{lr|cccc} \hline\hline
		&  & \multicolumn{4}{c}{$E_g$ } \\ \cline{3-6} 
		Topic & $N\quad$  & Mean     & Sd       & Min      & Max    \\
		\hline
		Pop art            & 4,013 & 5.506 & 1.000   & 0.893 & 7.520    \\
		Magic realism     & 2,401 & 3.534 & 1.122 & 1.765 & 6.988\\
		Art informel        & 1,874 & 5.304 & 0.851 & 1.500 & 7.336\\
		Tachisme           & 1,263 & 5.648 & 0.912 & 1.972 & 7.490\\
		Expressionism      & 1,228 & 5.195 & 1.040   & 1.617 & 7.673 \\
		Surrealism          & 977  & 5.319 & 1.033 & 1.972 & 7.490\\
		Impressionism       & 756  & 4.840 & 1.088 & 2.053 & 7.561\\
		Na{\"i}ve art (Primitivism)  & 693  & 5.450 & 0.977 & 1.446 & 7.465 \\
		Cubism             & 620  & 5.380 & 0.914 & 1.313 & 7.507\\
		Abstract expressionism   & 594  & 5.464 & 1.110 & 1.349   & 7.353 \\
		\textbf{Whole sample}           & 21,027 & 5.038 & 1.256 & 0.497 & 7.691\\	\hline\hline
	&&&&&\\
	%\begin{tablenotes}
		\multicolumn{6}{l}{\footnotesize{\emph{Note: }Table~\ref{tab:style1} presents the descriptive statistics of the SVD entropies of}}\\
		\multicolumn{6}{l}{grayscale values for top ten style categories as well as the whole sample. }  
	%\end{tablenotes}
\end{tabular}
\end{table}
%Because WikiArt use granular style categories, some styles in our sample do not have enough observations to run a subsample regression. Therefore, we only implement the overall hedonic regression by controlling for the style dummies. 

Limited by the sample size, we will not run the subsample regression (i.e., style by style) as in the robustness analysis of topic. We only run the aggregate regression by introducing the new control of style dummy $Style_{i}$ as follows: 
\begin{equation} \label{eq:regstyle}
	\log \, p_{i} = \alpha E_i + \sum_{k=1}^K\beta_{k}X_{ki} + 	\sum_{t=1}^T\gamma_{t}D_{it}+\eta\, Style_{i}+\epsilon_{i}, 
\end{equation} 
Table~\ref{tab:style} presents the regression outputs. The results (Specificatons (2) and (3) in Table~\ref{tab:style}) are mostly similar to what we obtained before.  

We also want to further control the topic for the above regression. So, we apply the previous topic classification methodology to this sample with style labels. We find that 69\% of this sample is either Untitled or Unknown, and the rest can then be classified into the regular ten topic categories.\footnote{See Appendix~\ref{appx c} for the details.} Controlling both the topic and the style of a painting (Specifications (4) and (5) in Table~\ref{tab:style}) does not change the result either: the SVD entropy still positively affects the sales price of the painting at 1\% significance level.

\begin{table}[htbp] \footnotesize \centering
\caption{The hedonic regressions with style dummy. \label{tab:style}}
	\begin{tabular}{l*{5}{c}}
	\hline\hline
	&\multicolumn{1}{c}{(1)}&\multicolumn{1}{c}{(2)}&\multicolumn{1}{c}{(3)}&\multicolumn{1}{c}{(4)}&\multicolumn{1}{c}{(5)}\\

	\hline
	$E_g$          &                     &       0.073\sym{***}&       0.149\sym{***}&       0.073\sym{***}&       0.144\sym{***}\\
	&                     &     (0.011)         &     (0.054)         &     (0.011)         &     (0.053)         \\
	[1em]
	$E_g^2$         &                     &                     &      -0.008         &                     &      -0.008         \\
	&                     &                     &     (0.006)         &                     &     (0.006)         \\
	[1em]
	Height      &       0.004\sym{*}  &       0.004\sym{*}  &       0.004\sym{*}  &       0.004\sym{*}  &       0.004\sym{*}  \\
	&     (0.002)         &     (0.002)         &     (0.002)         &     (0.002)         &     (0.002)         \\
	[1em]
	Height$^2$     &       0.000\sym{***}&       0.000\sym{***}&       0.000\sym{***}&       0.000\sym{***}&       0.000\sym{***}\\
	&     (0.000)         &     (0.000)         &     (0.000)         &     (0.000)         &     (0.000)         \\
	[1em]
	Width       &       0.012\sym{***}&       0.011\sym{***}&       0.011\sym{***}&       0.011\sym{***}&       0.012\sym{***}\\
	&     (0.002)         &     (0.002)         &     (0.002)         &     (0.002)         &     (0.002)         \\
	[1em]
	Width$^2$      &      -0.000\sym{***}&      -0.000\sym{***}&      -0.000\sym{***}&      -0.000\sym{***}&      -0.000\sym{***}\\
	&     (0.000)         &     (0.000)         &     (0.000)         &     (0.000)         &     (0.000)         \\
	[1em]
	Signature dummy  &       0.317\sym{***}&       0.315\sym{***}&       0.315\sym{***}&       0.313\sym{***}&       0.313\sym{***}\\
	&     (0.024)         &     (0.024)         &     (0.024)         &     (0.024)         &     (0.024)         \\
	[1em]
	Dated dummy &       0.330\sym{***}&       0.326\sym{***}&       0.326\sym{***}&       0.327\sym{***}&       0.327\sym{***}\\
	&     (0.023)         &     (0.023)         &     (0.023)         &     (0.023)         &     (0.023)         \\
	[1em]
	Topic dummy&  -    &   -   &  -    &     control     &  control       \\
   	[1em]
	\textit{\textbf{Style dummy}}&  \textit{\textbf{control}}    &   \textit{\textbf{control}}    &  \textit{\textbf{control}}     &     \textit{\textbf{control}}     &  \textit{\textbf{control}}       \\
	[1em]
	Artist dummy   &  control   & control     &   control   &   control     &  control      \\
	[1em]
	Medium dummy&   control   & control     &   control   &   control   &     control        \\
	[1em]
	Auction house dummy&   control   &  control    &   control   &  control     &  control        \\
	[1em]
	Auction city dummy&  control   & control    &   control    &   control     &  control         \\
	[1em]
	Year dummy&  control     &  control    &  control    &     control       &  control         \\
	[1em]
	Month dummy&  control    &   control    &  control     &     control     &  control        \\
	[1em]
	\hline
	\(N\)       &       21,027         &       21,027         &       21,027         &       21,027         &       21,027         \\
	adj. \(R^{2}\)&       0.759         &       0.759         &       0.759         &       0.760         &       0.760         \\
	\hline\hline
	&&&&&\\

%\begin{tablenotes}\\
	\multicolumn{6}{l}{\footnotesize{\emph{Note: }1.  Table~\ref{tab:style} presents the results of hedonic regressions of artworks' log prices}}\\
	\multicolumn{6}{l}{\footnotesize{on the information quantity measurements controlling for the style and other }}\\
	\multicolumn{6}{l}{\footnotesize{variables. The classification of style dummies is from \url{wikiart.org}.}} \\
	\multicolumn{6}{l}{\footnotesize{2.  The number of observations ($N$) and the adjusted R-squared (Adj. $R^2$) are }}\\
	\multicolumn{6}{l}{\footnotesize{presented at the bottom of the table.}}\\
	\multicolumn{6}{l}{\footnotesize{3.  Standard errors are in parentheses}}\\
	\multicolumn{6}{l}{\footnotesize{4. ***, **, and * denote significance at the 1\%, 5\%, and 10\% levels, respectively.}}
%\end{tablenotes}
\end{tabular}
\end{table}

\section{Discussion on the application} \label{sec:application}

So far, we demonstrate that $E_g$, the SVD entropy that measures the amount of backbone information of a painting, can significantly positively affect the sales price of the painting in a robust way. Nowadays, it is very convenient to obtian a high-quality digital image for almost any painting. So, creating some kind of measurements based on the pixel-level information will surely find more and more applications in art pricing research. The most common measurements in literature are color variables. However, we show below that the color attributes of a painting, although being objective measurements too, may not be robust regressors for the sales price. 

The literature generally finds that the percentage of blue or red in a painting can fetch a price premium based on various testing samples (e.g., the Picasso sample in \citet{stepanova2019impact} and the shape/pattern-controlled sample in \citet{ma2022colors}). In this paper, we construct similar color measurements, i.e., the percentages of blue and red (denoted as Redpct and Bluepct), and test their effects on the painting's sales price based on the matched sample with style labels.\footnote{We define the percentage of red as the proportion of pixels with hue values in the range of 0-14 or 165-179, and the percentage of blue as the proportion of pixels with hue values in the range of 105-134.} The results in Table~\ref{tab:color} show that the percentage of red is not significant while the percentage of blue even has a negative effect on the sales price. Furthermore, neither blue nor red color percentage interacts with $E_g$. It implies that the color compositions cannot affect price via the Shannon information channel. So, if we can extract the Shannon information from colors, it must be derived from the colors' attachments to the painting's structural elements like line, shape, form, etc., which we define as backbone information before. In that case, removing colors from an image may not lose the image's Shannon information.

\begin{table}[htbp] \scriptsize\centering
	\caption{The hedonic regressions with colors. \label{tab:color}}
	\begin{tabular}{l*{5}{c}}
		\hline\hline
		&\multicolumn{1}{c}{(1)}&\multicolumn{1}{c}{(2)}&\multicolumn{1}{c}{(3)}&\multicolumn{1}{c}{(4)}&\multicolumn{1}{c}{(5)}\\
		
		\hline
		$E_g$           &                     &       0.075\sym{***}&              0.074\sym{***}&       0.075\sym{***}&       0.075\sym{***}\\
		&                     &     (0.011)               &     (0.014)         &     (0.011)         &     (0.014)         \\
		[1em]
		Redpct      &       0.035         &       0.022         &              0.050         &       0.029         &       0.075         \\
		&     (0.043)         &     (0.043)                  &     (0.159)         &     (0.043)         &     (0.159)         \\
		[1em]
		Bluepct      &      -0.090         &      -0.128\sym{**} &           -0.282         &      -0.119\sym{*}  &      -0.262  \\
		&     (0.063)         &     (0.063)                 &     (0.253)         &     (0.063)         &     (0.063)         \\
		[1em]
		$E_g$*\hspace{0.03cm}Redpct      &                                          &                     &      -0.006         &                     &              -0.009       \\
		&                     &                                         &     (0.031)         &                     &        (0.031)             \\
		[1em]
		$E_g$*\hspace{0.03cm}Bluepct     &                                          &                     &       0.030         &                     &         0.028            \\
		&                     &                                         &     (0.048)         &                     &        (0.048)             \\
		[1em]
		Height      &       0.004\sym{*}  &       0.004\sym{*}  &              0.004\sym{*}  &       0.004\sym{*}  &       0.004\sym{*}  \\
		&     (0.002)         &     (0.002)               &     (0.002)         &     (0.002)         &     (0.002)         \\
		[1em]
		Height$^2$     &       0.000\sym{***}&             0.000\sym{***}&       0.000\sym{***}&       0.000\sym{***}&       0.000\sym{***}\\
		&     (0.000)         &     (0.000)                  &     (0.000)         &     (0.000)         &     (0.000)         \\
		[1em]
		Width       &       0.012\sym{***}&              0.011\sym{***}&       0.011\sym{***}&       0.011\sym{***}&       0.011\sym{***}\\
		&     (0.002)         &     (0.002)                 &     (0.002)         &     (0.002)         &     (0.002)         \\
		[1em]
		Width$^2$     &      -0.000\sym{***}&         -0.000\sym{***}&      -0.000\sym{***}&      -0.000\sym{***}&      -0.000\sym{***}\\
		&     (0.000)         &     (0.000)                 &     (0.000)         &     (0.000)         &     (0.000)         \\
		[1em]
		Signature dummy  &       0.318\sym{***}&             0.315\sym{***}&       0.315\sym{***}&       0.313\sym{***}&       0.313\sym{***}\\
		&     (0.024)         &     (0.024)                 &     (0.024)         &     (0.024)         &     (0.024)         \\
		[1em]
		Dated dummy  &       0.329\sym{***}&       0.324\sym{***}&             0.324\sym{***}&       0.325\sym{***}&       0.325\sym{***}\\
		&     (0.023)         &     (0.023)                 &     (0.023)         &     (0.023)         &     (0.023)         \\
		[1em]
		Topic dummy&  - &  -       &  -     &     control    &  control      \\
		[1em]
		Style dummy&  control    &   control        &     control    &  control &  control       \\
		[1em]
		Artist dummy   &  control   & control       &   control     &  control  &  control      \\
		[1em]
		Medium dummy&   control   & control       &   control   &     control   &  control       \\
		[1em]
		Auction house dummy&   control     &   control   &  control     &  control    &  control      \\
		[1em]
		Auction city dummy&  control      &   control    &   control     &  control    &  control       \\
		[1em]
		Year dummy&  control     &  control        &     control       &  control    &  control       \\
		[1em]
		Month dummy&  control    &   control        &     control     &  control     &  control     \\
		[1em]
		\hline
		\hline
		\(N\)        &       21,027                &       21,027         &       21,027         &       21,027         &       21,027         \\
		adj. \(R^{2}\)&       0.759                &       0.759         &       0.759         &       0.760         &       0.760         \\
		\hline\hline
		&&&&&\\

	%\begin{tablenotes}\\
	\multicolumn{6}{l}{\footnotesize{\emph{Note: }1.  Table~\ref{tab:color} presents the results of hedonic regressions of artworks' log prices }}\\
	\multicolumn{6}{l}{on the information quantity measurements and the color attributes as well as other control}\\
	\multicolumn{6}{l}{variables.}\\
	\multicolumn{6}{l}{2.  The number of observations ($N$) and the adjusted R-squared (Adj. $R^2$) are presented at the}\\ 
	\multicolumn{6}{l}{ bottom of the table.}\\
		\multicolumn{6}{l}{3.  Standard errors are in parentheses}\\
		\multicolumn{6}{l}{4. ***, **, and * denote significance at the 1\%, 5\%, and 10\% levels, respectively.}
	%\end{tablenotes}
		\end{tabular}
\end{table}

Our testing sample can be considered a random sample as it is obtained from the matching of two different databases. So we can say that compared to the color attributes, the SVD entropy is more robust against the sample selection. Moreover, we will show below that the SVD entropy also improves the model fit more than the other commonly adopted variables that describe a painting's content. 

These common content variables in the hedonic regression model include: siganiture dummy, dated dummy, topic dummy and style dummy. For the model fit comparison here, we will not consider the topic and style dummies for two reasons. First, different sources have different classification criteria, so they appear to be subjective relative to the other content variables. Second, they generally contain quite a few categories, which give them a natural advantage in increasing the R squared. So we will only focus on the three objective measurements concerning the content, i.e., the SVD entropy, the signature dummy and the dated dummy.    

We take two different approaches to compare each variable's contribution to the adjusted R squared based on the largest sample (529,654 observations) used in the benchmark regression. The first approach is to control all the other variables while adding these three content variables one by one to see which improves the adjusted R squared the most. The results are presented in the upper part of Table~\ref{tab:adjr2}. The second approach is to run the regression with the full set of variables and then omit these three variables alternatively to see which decreases the adjusted R squared the least. The lower part of Table~\ref{tab:adjr2} shows the results. Both approaches indicate that the SVD entropy $E_g$ enhances the model fit more than the signature dummy as well as the dated dummy. Note that the improvement brought by each variable to the model's R squared appears small. This is because the regression includes more than three thousand regressors, so each single variable's contribution is diminished.   

In sum, considering the SVD entropy's convenient data accessibility, straightforward calculation algorithm, as well as its advantages in variable significance, sample robustness and model fit, we expect its wide application in future research.

 \begin{table}[htbp]\centering\small
	\def\sym#1{\ifmmode^{#1}\else\(^{#1}\)\fi}
	\caption{Model fit (Adjusted \(R^{2}\)) comparisons \label{tab:adjr2}}
	\begin{tabular}{l*{4}{c}}
		\hline\hline
		&\multicolumn{1}{c}{(1)}&\multicolumn{1}{c}{(2)}&\multicolumn{1}{c}{(3)}&\multicolumn{1}{c}{(4)}\\
		\hline
		$E_g$          &                     &       0.104\sym{***}&                     &                     \\
		&                     &     (0.002)         &                     &                     \\
		[1em]
		Signature dummy  &                     &                     &       0.164\sym{***}&                     \\
		&                     &                     &     (0.004)         &                     \\
		[1em]
		Dated dummy &                     &                     &                     &       0.200\sym{***}\\
		&                     &                     &                     &     (0.004)         \\
		All other control variables   &  control   & control     &   control   &   control \\
		[1em]
		\hline
		\(N\)       &      529,654         &      529,654         &      529,654         &      529,654         \\
		Adj. \(R^{2}\)&       0.655         &       0.657         &       0.656         &       0.657         \\
		\hline\hline
		\vspace{0.5cm}
		&&&&\\
		\hline\hline		
		&\multicolumn{1}{c}{(5)}&\multicolumn{1}{c}{(6)}&\multicolumn{1}{c}{(7)}&\multicolumn{1}{c}{(8)}\\
		\hline
		$E_g$          &       0.102\sym{***}&                     &       0.102\sym{***}&       0.104\sym{***}\\
		&     (0.002)         &                     &     (0.002)         &     (0.002)         \\
		[1em]
		Signature dummy  &       0.139\sym{***}&       0.140\sym{***}&                     &       0.162\sym{***}\\
		&     (0.004)         &     (0.004)         &                     &     (0.004)         \\
		[1em]
		Dated dummy &       0.176\sym{***}&       0.181\sym{***}&       0.195\sym{***}&                     \\
		&     (0.004)         &     (0.004)         &     (0.004)         &                     \\
		All other control varialbes   &  control   & control     &   control   &   control \\
		[1em]
		\hline
		\(N\)       &      529,654         &      529,654         &      529,654         &      529,654         \\
		Adj. \(R^{2}\)&       0.659         &       0.657         &       0.659         &       0.658         \\
		\hline\hline
		&&&&\\
	%\begin{tablenotes}\\
	\multicolumn{5}{l}{\footnotesize{\emph{Note: }1.  Table~\ref{tab:adjr2} compares the contributions to the model fit of three content variables:}}\\
	 \multicolumn{5}{l}{\footnotesize{SVD entropy, signature dummy and dated dummy.}} \\
		\multicolumn{5}{l}{\footnotesize{2.  The number of observations ($N$) and the adjusted R-squared (Adj. $R^2$) are presented }}\\
			\multicolumn{5}{l}{\footnotesize{at the bottom of the two parts of this table.}}\\
		\multicolumn{5}{l}{\footnotesize{3.  Standard errors are in parentheses}}\\
		\multicolumn{5}{l}{\footnotesize{4. ***, **, and * denote significance at the 1\%, 5\%, and 10\% levels, respectively.}}
	%\end{tablenotes}
	
\end{tabular}
\end{table}

\section{Conclusion} \label{sec:conc}

The hedonic models of art pricing often lack fine measurements of painting content. The traditional content measurements are normally categorical data, e.g, dummies of signature, dated, topic, style, and sometimes expert assessment. This paper proposes a new measurement to account for content heterogeneity, which is neither categorical nor subjective, i.e., the Shannon information quantity of the painting. We suggest using SVD entropy to measure the Shannon information quantity based on each painting's digital image. Our empirical tests show that the Shannon information quantity (measured by SVD entropy) of a painting is a significant and robust pricing factor. It is also superior to the other usual content variables, such as the signature dummy and the dated dummy, in terms of the improvement to model fit. 

Our research demonstrates that applying computer graphic techniques to art pricing is valuable. For future studies, we may test the effect of SVD entropy with other available data samples to assess its robustness. Also, we can investigate the information in three-dimensional artworks, such as sculptures and porcelains, so as to improve their pricing models. 

\clearpage
% Remove or comment out the next two lines if you are not using bibtex.
\bibliographystyle{elsarticle-harv}
\begin{singlespace}
\bibliography{ArtValue}
\end{singlespace}

% The appendix command is issued once, prior to all appendices, if any.
\newpage
\appendix
\section{Grayscale and HSV}
\label{appx a}
The grayscale determines the relative distance (or grayness) between white and black. The grayscale is characterized by the three-dimensional RGB system. Let $R$, $G$, and $B$ be the three proportion numbers in the RGB system, and the grayscale of a pixel is given by the following formula: 
$$ Grayscale = 0.3 R + 0.59 G + 0.11 B $$

REG also determines the hue of a pixel. It is a one-dimensional degree number that is reduced from the RGB to specifically represent the unique color of a given pixel. The hue is defined to range from $ 0^{\circ} $ to $ 360^{\circ} $, starting from red and moving in an anti-clockwise direction. Let max and min be the maximum and minimum of $R$, $G$, and $B$, then the hue of a pixel can be defined as:

$$\quad \textit{Hue}=
\begin{cases}
	\text{undefined},& \text{if } max=min\\
	60^{\circ}\times \dfrac{G-B}{max-min} + 0^{\circ}& \text{if } max=R \text{ and } G\geq B\\
	60^{\circ}\times \dfrac{G-B}{max-min} + 360^{\circ}& \text{if }max=R \text{ and }G\textless B\\
	60^{\circ}\times \dfrac{B-R}{max-min} + 120^{\circ}& \text{if }max=G \\
	60^{\circ}\times \dfrac{R-G}{max-min} + 240^{\circ}& \text{if }max=B 
\end{cases}$$

\vspace{0.3cm}

The value measures the brightness of the color. Usually, the value ranges from 0\% (black) to 100\% (white).
\vspace{0.03cm}
\begin{center}
	$\textit{Value}=max(R,G,B)$
\end{center}

\vspace{0.2cm}
A color can be seen as the result of mixing a spectral color with white. Then the saturation indicates how close the color is to the spectral color. 
\vspace{0.07cm}
$$\quad \textit{Saturation}=
\begin{cases}
	\dfrac{max(R,G,B)-min(R,G,B),}{max(R,G,B)}& \text{if } max(R,G,B)\neq 0\\
	0 & \text{otherwise }\\
	\end{cases}$$

%\begin{figure}[ht!]
%   \caption{Hue.
%		\label{f:hue}}
%	\centering
	
%	\includegraphics[width=0.4\textwidth]{huebar.png}\hspace{2em}
	
%	\begin{figurenotes}
%		Figure~\ref{f:hue} plots the hue representation of colors.
%	\end{figurenotes}
%\end{figure}

\medskip
\medskip
\section{Detailed Outputs of Table~\ref{tab:benchmark}}
\label{appx b}
We provide the more detailed version of Table~\ref{tab:benchmark} as follows.
\begin{table}[htbp]\footnotesize
	\centering
	\caption{Benchmark regressions with more details.}
	\begin{tabular}{l*{5}{c}}
		\hline\hline
		&\multicolumn{1}{c}{(1)}&\multicolumn{1}{c}{(2)}&\multicolumn{1}{c}{(3)}&\multicolumn{1}{c}{(4)}&\multicolumn{1}{c}{(5)}\\
		\hline
		$E_G$          &                     &       0.102\sym{***}&       0.021\sym{**}  &       0.110\sym{***}&       0.107         \\
		&                     &     (0.002)         &     (0.010)         &     (0.002)         &     (0.002)         \\
		[1em]
		$E_G^2$          &                     &                     &       0.008\sym{***}&                     &      \\
		&                     &                     &     (0.001)         &                     &              \\
		[1em]
		Height      &       0.006\sym{***}&       0.006\sym{***}&       0.007\sym{***}&       0.009\sym{***}&       0.006\sym{***}\\
		&     (0.000)         &     (0.000)         &     (0.000)         &     (0.000)         &     (0.000)         \\
		[1em]
		Height$^2$      &      -0.000\sym{***}&      -0.000\sym{***}&      -0.000\sym{***}&      -0.000\sym{***}&      -0.000\sym{***}\\
		&     (0.000)         &     (0.000)         &     (0.000)         &     (0.000)         &     (0.000)         \\
		[1em]
		$E_g$*\hspace{0.05cm}Height   &                     &                     &                     &      -0.000\sym{***}&                     \\
		&                     &                     &                     &     (0.000)         &                     \\
		[1em]
		Width       &       0.006\sym{***}&       0.006\sym{***}&       0.006\sym{***}&       0.006\sym{***}&       0.007\sym{***}\\
		&     (0.000)         &     (0.000)         &     (0.000)         &     (0.000)         &     (0.000)         \\
		[1em]
		Width$^2$     &      -0.000\sym{***}&      -0.000\sym{***}&      -0.000\sym{***}&      -0.000\sym{***}&      -0.000\sym{***}\\
		&     (0.000)         &     (0.000)         &     (0.000)         &     (0.000)         &     (0.000)         \\
		[1em]
		$E_g$*\hspace{0.05cm}Width    &                     &                     &                     &                     &      -0.000\sym{***}\\
		&                     &                     &                     &                     &     (0.000)         \\
		[1em]
		Signature dummy  &       0.140\sym{***}&       0.139\sym{***}&       0.139\sym{***}&       0.139\sym{***}&       0.139\sym{***}\\
		&     (0.004)         &     (0.004)         &     (0.004)         &     (0.004)         &     (0.004)         \\
		[1em]
		Dated dummy &       0.181\sym{***}&       0.176\sym{***}&       0.176\sym{***}&       0.176\sym{***}&       0.176\sym{***}\\
		&     (0.004)         &     (0.004)         &     (0.004)         &     (0.004)         &     (0.004)         \\
		[1em]
		\hline
		\textbf{Artist}&&&&&\\
		[1em]
		Pablo Ruiz Picasso% \_Iartist\_2263
		&      -0.072         &      -0.019         &      -0.019         &      -0.017         &      -0.018         \\
		&     (0.048)         &     (0.048)         &     (0.048)         &     (0.048)         &     (0.048)         \\
		[1em]
		Marc Chagall% \_Iartist\_1975
		&      -0.289\sym{***}&      -0.291\sym{***}&      -0.291\sym{***}&      -0.290\sym{***}&      -0.291\sym{***}\\
		&     (0.049)         &     (0.048)         &     (0.048)         &     (0.048)         &     (0.048)         \\
		[1em]
		Andy Warhol% \_Iartist\_192
		&       0.058         &       0.091\sym{*}         &       0.092\sym{*}         &       0.092\sym{*}         &       0.092\sym{*}         \\
		&     (0.049)         &     (0.048)         &     (0.048)         &     (0.048)         &     (0.048)         \\
		[1em]
		Joan Mir{\'o} %\_Iartist\_1595
		&      -0.598\sym{***}&      -0.576\sym{***}&      -0.572\sym{***}&      -0.574\sym{***}&      -0.575\sym{***}\\
		&     (0.049)         &     (0.049)         &     (0.049)         &     (0.049)         &     (0.049)         \\
		[1em]
		Salvador Dal{\'i}%(1904-1989) Spain  \_Iartist\_2607
		&      -1.175\sym{***}&      -1.082\sym{***}&      -1.079\sym{***}&      -1.081\sym{***}&      -1.081\sym{***}\\
		&     (0.049)         &     (0.049)         &     (0.049)         &     (0.049)         &     (0.049)         \\
		[1em]		
		Rembrandt Harmensz Van Rijn %\_Iartist\_2450
		&      -0.109\sym{**}  &      -0.134\sym{***} &      -0.137\sym{***} &      -0.135\sym{**}  &      -0.116\sym{**}  \\
		&     (0.051)         &     (0.051)         &     (0.051)         &     (0.051)         &     (0.051)         \\
		[1em]
		Victor Vasarely  %\_Iartist\_2798
		&      -1.631\sym{***}&      -1.557\sym{***}&      -1.553\sym{***}&      -1.557\sym{***}&      -1.557\sym{***}\\
		&     (0.051)         &     (0.051)         &     (0.051)         &     (0.051)         &     (0.051)         \\
		[1em]
		Alexander ``Sandy'' Calder %\_Iartist\_96 
		&      -0.639\sym{***}&      -0.624\sym{***}&      -0.622\sym{***}&      -0.622\sym{***}&      -0.622\sym{***}\\
		&     (0.054)         &     (0.054)         &     (0.054)         &     (0.054)         &     (0.054)         \\
		[1em]
	
	\end{tabular}
\end{table}

\begin{table}[htbp]\footnotesize
	\centering
	%	\caption{Benchmark models.}
	\begin{tabular}{l*{5}{c}}
		%		\hline
		%		&\multicolumn{1}{c}{(1)}&\multicolumn{1}{c}{(2)}&\multicolumn{1}{c}{(3)}&\multicolumn{1}{c}{(4)}&\multicolumn{1}{c}{(5)}\\
		%
			David Hockney %(1937-) England	\_Iartist\_599
		&      -0.737\sym{***}&      -0.637\sym{***}&      -0.638\sym{***}&      -0.636\sym{***}&      -0.636\sym{***}\\
		&     (0.054)         &     (0.054)         &     (0.054)         &     (0.054)         &     (0.054)         \\
		[1em]	
		Albrecht D{\"u}rer %(1471-1528) Germany  \_Iartist\_89 
		&      -0.314\sym{***}&      -0.369\sym{***}&      -0.379\sym{***}&      -0.371\sym{***}&      -0.371\sym{***}\\
		&     (0.055)         &     (0.055)         &     (0.055)         &     (0.055)         &     (0.055)         \\
		[1em]
		Other artists omitted  &&&&&\\		
		\hline
		\textbf{Medium}&&&&&\\
		[1em]
		Oil on canvas  %\_Iadj\_mediu\_26
		&       0.220\sym{***}&       0.210\sym{***}&       0.212\sym{***}&       0.211\sym{***}&       0.211\sym{***}\\
		&     (0.019)         &     (0.019)         &     (0.019)         &     (0.019)         &     (0.019)         \\
		[1em]
		Lithograph %\_Iadj\_mediu\_18
		&      -2.548\sym{***}&      -2.537\sym{***}&      -2.534\sym{***}&      -2.537\sym{***}&      -2.536\sym{***}\\
		&     (0.020)         &     (0.020)         &     (0.020)         &     (0.020)         &     (0.020)         \\
		[1em]
		Etching  %\_Iadj\_mediu\_10
		&      -2.167\sym{***}&      -2.141\sym{***}&      -2.140\sym{***}&      -2.141\sym{***}&      -2.140\sym{***}\\
		&     (0.021)         &     (0.021)         &     (0.021)         &     (0.021)         &     (0.021)         \\
		[1em]
		Color lithograph %\_Iadj\_mediu\_5
		&      -2.432\sym{***}&      -2.415\sym{***}&      -2.411\sym{***}&      -2.415\sym{***}&      -2.414\sym{***}\\
		&     (0.021)         &     (0.021)         &     (0.021)         &     (0.021)         &     (0.021)         \\
		[1em]
		Watercolor  %\_Iadj\_mediu\_49
		&      -1.049\sym{***}&      -1.013\sym{***}&      -1.009\sym{***}&      -1.012\sym{***}&      -1.012\sym{***}\\
		&     (0.021)         &     (0.021)         &     (0.021)         &     (0.021)         &     (0.021)         \\
		[1em]
		Oil on board  %\_Iadj\_mediu\_25
		&      -0.256\sym{***}&      -0.262\sym{***}&      -0.260\sym{***}&      -0.262\sym{***}&      -0.262\sym{***}\\
		&     (0.022)         &     (0.022)         &     (0.022)         &     (0.022)         &     (0.022)         \\
		[1em]
		Oil on panel  %\_Iadj\_mediu\_29
		&      -0.180\sym{***}&      -0.191\sym{***}&      -0.189\sym{***}&      -0.191\sym{***}&      -0.191\sym{***}\\
		&     (0.022)         &     (0.022)         &     (0.022)         &     (0.022)         &     (0.022)         \\
		[1em]
		Gelatin silver print  %\_Iadj\_mediu\_13
		&      -1.611\sym{***}&      -1.600\sym{***}&      -1.597\sym{***}&      -1.600\sym{***}&      -1.599\sym{***}\\
		&     (0.026)         &     (0.026)         &     (0.026)         &     (0.026)         &     (0.026)         \\
		[1em]
		Print  %\_Iadj\_mediu\_42
		&      -1.947\sym{***}&      -1.933\sym{***}&      -1.930\sym{***}&      -1.932\sym{***}&      -1.932\sym{***}\\
		&     (0.023)         &     (0.023)         &     (0.023)         &     (0.023)         &     (0.023)         \\
		[1em]
		Watercolor on paper %	\_Iadj\_mediu\_50
		&      -0.938\sym{***}&      -0.907\sym{***}&      -0.903\sym{***}&      -0.906\sym{***}&      -0.906\sym{***}\\
		&     (0.023)         &     (0.023)         &     (0.023)         &     (0.023)         &     (0.023)         \\
		[1em]
		Other media omitted  &&&&&\\
		\hline
		\textbf{Auction house}&&&&&\\
		[1em]
		Christies %company_10
		&       1.036\sym{***}&       1.072\sym{***}&       1.076\sym{***}&       1.071\sym{***}&       1.071\sym{***}\\
		&     (0.020)         &     (0.019)         &     (0.019)         &     (0.019)         &     (0.019)         \\
		[1em]
		Bonhams  %company_5
		&       0.081\sym{***}&       0.061\sym{***} &       0.060\sym{***} &       0.060\sym{***} &       0.061\sym{***} \\
		&     (0.020)         &     (0.019)         &     (0.019)         &     (0.019)         &     (0.019)         \\
		[1em]
		Swann galleries  %company_21
		&       0.006         &      -0.012         &      -0.010         &      -0.012         &      -0.012         \\
		&     (0.022)         &     (0.022)         &     (0.022)         &     (0.022)         &     (0.022)         \\
		[1em]
		Bruun rasmussen  %company_7
		&      -0.273\sym{***}&      -0.316\sym{***}&      -0.321\sym{***}&      -0.315\sym{***}&      -0.316\sym{***}\\
		&     (0.030)         &     (0.029)         &     (0.029)         &     (0.029)         &     (0.029)         \\
		[1em]
		Artcurial - briest-poulain-le fur  %company_2
		&       0.219\sym{***}&       0.217\sym{***}&       0.218\sym{***}&       0.216\sym{***}&       0.217\sym{***}\\
		&     (0.022)         &     (0.022)         &     (0.022)         &     (0.022)         &     (0.022)         \\
		[1em] 
		Phillips  %company_18
		&       0.867\sym{***}&       0.858\sym{***}&       0.860\sym{***}&       0.858\sym{***}&       0.858\sym{***}\\
		&     (0.023)         &     (0.023)         &     (0.023)         &     (0.023)         &     (0.023)         \\
		[1em]
	
	\end{tabular}
\end{table}

\begin{table}[htbp]\footnotesize
	\centering
	%\caption{Benchmark models.}
	\begin{tabular}{l*{5}{c}}
		%\hline
		%&\multicolumn{1}{c}{(1)}&\multicolumn{1}{c}{(2)}&\multicolumn{1}{c}{(3)}&\multicolumn{1}{c}{(4)}&\multicolumn{1}{c}{(5)}\\
		%
		%\hline
			Dorotheum %company_11
		&      -0.197\sym{***}&      -0.163\sym{***}&      -0.158\sym{***}&      -0.163\sym{***}&      -0.163\sym{***}\\
		&     (0.032)         &     (0.032)         &     (0.032)         &     (0.032)         &     (0.032)         \\
		[1em]
		Bukowskis  %company_8
		&       0.805\sym{***}&       0.757\sym{***}&       0.754\sym{***}&       0.757\sym{***}&       0.757\sym{***}\\
		&     (0.018)         &     (0.018)         &     (0.018)         &     (0.018)         &     (0.018)         \\	
		[1em]
		Auktionshuset metropol ab %company_3
		&      -0.589\sym{***}&      -0.571\sym{***}&      -0.569\sym{***}&      -0.571\sym{***}&      -0.571\sym{***}\\
		&     (0.018)         &     (0.018)         &     (0.018)         &     (0.018)         &     (0.018)         \\
		[1em]
		Millon \& associes  %company_16
		&      -0.145\sym{***}&      -0.125\sym{***}&      -0.121\sym{***}&      -0.125\sym{***}&      -0.125\sym{***}\\
		&     (0.028)         &     (0.028)         &     (0.028)         &     (0.028)         &     (0.028)         \\
		[1em]
		Sothebys %company_20
		&       1.244\sym{***}&       1.249\sym{***}&       1.251\sym{***}&       1.248\sym{***}&       1.248\sym{***}\\
		&     (0.024)         &     (0.024)         &     (0.024)         &     (0.024)         &     (0.024)         \\
		[1em]
		Other houses omitted  &&&&&\\
		\hline
		\textbf{Auction city}&&&&&\\
		[1em]
		New York  %city_13
		&       0.647\sym{***}&       0.655\sym{***}&       0.655\sym{***}&       0.655\sym{***}&       0.655\sym{***}\\
		&     (0.020)         &     (0.020)         &     (0.020)         &     (0.020)         &     (0.020)         \\
		[1em]
		London	%city_11
		&       0.501\sym{***}&       0.503\sym{***}&       0.503\sym{***}&       0.503\sym{***}&       0.503\sym{***}\\
		&     (0.019)         &     (0.019)         &     (0.019)         &     (0.019)         &     (0.019)         \\
		[1em]
		Paris  %city_15
		&       0.497\sym{***}&       0.505\sym{***}&       0.505\sym{***}&       0.505\sym{***}&       0.505\sym{***}\\
		&     (0.020)         &     (0.020)         &     (0.020)         &     (0.020)         &     (0.020)         \\
		[1em]
		Stockholm  %city_16
		&       0.041         &       0.030         &       0.032         &       0.030         &       0.030        \\
		&     (0.025)         &     (0.025)         &     (0.025)         &     (0.025)         &     (0.025)         \\
		[1em]
		Berlin %city_2
		&       0.496\sym{***}&       0.466\sym{***}&       0.464\sym{***}&       0.465\sym{***}&       0.465\sym{***}\\
		&     (0.024)         &     (0.024)         &     (0.024)         &     (0.024)         &     (0.024)         \\
		[1em]
		Cologne %city_6
		&       0.851\sym{***}&       0.860\sym{***}&       0.860\sym{***}&       0.860\sym{***}&       0.860\sym{***}\\
		&     (0.024)         &     (0.024)         &     (0.024)         &     (0.024)         &     (0.024)         \\
		[1em]
		Vercelli %city_19
		&       0.083\sym{***} &       0.079\sym{**}  &       0.079\sym{**}  &       0.078\sym{**}  &       0.065\sym{**}  \\
		&     (0.032)         &     (0.032)         &     (0.032)         &     (0.032)         &     (0.032)         \\
		[1em]
		Bern %city_3
		&       0.799\sym{***}&       0.787\sym{***}&       0.786\sym{***}&       0.786\sym{***}&       0.787\sym{***}\\
		&     (0.025)         &     (0.024)         &     (0.024)         &     (0.024)         &     (0.024)         \\
		[1em]
		Havnen %city_10
		&      -0.408\sym{***}&      -0.404\sym{***}&      -0.403\sym{***}&      -0.406\sym{***}&      -0.419\sym{***}\\
		&     (0.030)         &     (0.030)         &     (0.030)         &     (0.030)         &     (0.030)         \\
		[1em]
		Dublin %city_7
		&       0.807\sym{***}&       0.801\sym{***}&       0.802\sym{***}&       0.801\sym{***}&       0.802\sym{***}\\
		&     (0.030)         &     (0.030)         &     (0.030)         &     (0.030)         &     (0.030)         \\
		[1em]
		Other cities omitted  &&&&&\\
		
		\hline
		\textbf{Year}&&&&&\\
		[1em]
		Year 2004 %year_2004 The same
		&       0.005         &       0.005         &       0.002         &       0.007         &      0.006         \\
		&     (0.475)         &     (0.474)         &     (0.474)         &     (0.474)         &     (0.474)         \\
		[1em]
		Year 2005 &       0.343         &       0.441         &       0.435         &       0.441         &       0.441         \\
		&     (0.440)         &     (0.439)         &     (0.439)         &     (0.439)         &     (0.439)         \\
		[1em]
		Year 2006 &       0.129         &       0.136         &       0.133         &       0.138         &       0.137         \\
		&     (0.440)         &     (0.438)         &     (0.438)         &     (0.438)         &     (0.438)         \\
		[1em]
		
	\end{tabular}
\end{table}

\begin{table}[htbp]\footnotesize
	\centering
	%\caption{Benchmark models.}
	\begin{tabular}{l*{5}{c}}
		%\hline
		%&\multicolumn{1}{c}{(1)}&\multicolumn{1}{c}{(2)}&\multicolumn{1}{c}{(3)}&\multicolumn{1}{c}{(4)}&\multicolumn{1}{c}{(5)}\\
		%
		%\hline
		Year 2007 &       0.201         &       0.194         &       0.190         &       0.197         &       0.196         \\
		&     (0.440)         &     (0.438)         &     (0.438)         &     (0.438)         &     (0.438)         \\		
		[1em]
		Year 2008 
		&       0.095         &       0.094         &       0.090         &       0.096         &       0.095         \\
		&     (0.440)         &     (0.438)         &     (0.438)         &     (0.438)         &     (0.438)         \\
		[1em]
		Year 2009 &       0.016         &       0.019         &       0.016         &       0.022         &       0.021        \\
		&     (0.440)         &     (0.438)         &     (0.438)         &     (0.438)         &     (0.438)         \\
		[1em]
		Year 2010 &       0.044         &       0.051         &       0.048         &       0.053         &       0.052         \\
		&     (0.440)         &     (0.438)         &     (0.438)         &     (0.438)         &     (0.438)         \\
		[1em]
		Year 2011 &       0.066         &       0.077         &       0.074         &       0.079         &       0.078         \\
		&     (0.440)         &     (0.438)         &     (0.438)         &     (0.438)         &     (0.438)         \\
		[1em]
		Year 2012 &       0.043         &       0.051         &       0.048         &       0.054         &       0.053         \\
		&     (0.440)         &     (0.438)         &     (0.438)         &     (0.438)         &     (0.438)         \\
		[1em]
		Year 2013 &       0.045         &       0.053         &       0.049         &       0.055         &       0.054         \\
		&     (0.440)         &     (0.438)         &     (0.438)         &     (0.438)         &     (0.438)         \\
		[1em]
		Year 2014 &      -0.018         &      -0.022         &      -0.026         &      -0.019         &      -0.020         \\
		&     (0.440)         &     (0.438)         &     (0.438)         &     (0.438)         &     (0.438)         \\
		[1em]
		Year 2015 &      -0.037         &      -0.044         &      -0.048         &      -0.042         &      -0.043         \\
		&     (0.440)         &     (0.438)         &     (0.438)         &     (0.438)         &     (0.438)         \\
		[1em]
		\hline
		\textbf{Month}&&&&&\\
		[1em]
		February  %month_2 following are same
		&       0.222\sym{***}&       0.213\sym{***}&       0.213\sym{***}&       0.213\sym{***}&       0.213\sym{***}\\
		&     (0.012)         &     (0.012)         &     (0.012)         &     (0.012)         &     (0.012)         \\
		[1em]
		March    &       0.145\sym{***}&       0.135\sym{***}&       0.135\sym{***}&       0.135\sym{***}&       0.135\sym{***}\\
		&     (0.011)         &     (0.011)         &     (0.011)         &     (0.011)         &     (0.011)         \\
		[1em]
		April   &       0.217\sym{***}&       0.207\sym{***}&       0.206\sym{***}&       0.207\sym{***}&       0.207\sym{***}\\
		&     (0.011)         &     (0.011)         &     (0.011)         &     (0.011)         &     (0.011)         \\
		[1em]
		May   &       0.403\sym{***}&       0.392\sym{***}&       0.392\sym{***}&       0.392\sym{***}&       0.392\sym{***}\\
		&     (0.011)         &     (0.011)         &     (0.011)         &     (0.011)         &     (0.011)         \\
		[1em]
		June   &       0.379\sym{***}&       0.373\sym{***}&       0.373\sym{***}&       0.373\sym{***}&       0.373\sym{***}\\
		&     (0.011)         &     (0.011)         &     (0.011)         &     (0.011)         &     (0.011)         \\
		[1em]
		July   &      -0.064\sym{***}&      -0.069\sym{***}&      -0.070\sym{***}&      -0.069\sym{***}&      -0.069\sym{***}\\
		&     (0.012)         &     (0.012)         &     (0.012)         &     (0.012)         &     (0.012)         \\
		[1em]
		August   &       0.029\sym{**}  &       0.022         &       0.021         &       0.022         &       0.022         \\
		&     (0.015)         &     (0.014)         &     (0.014)         &     (0.014)         &     (0.014)         \\
		[1em]
		September   &       0.002         &      -0.004         &      -0.004         &      -0.004         &      -0.004         \\
		&     (0.011)         &     (0.011)         &     (0.011)         &     (0.011)         &     (0.011)         \\
		[1em]
		October  &       0.213\sym{***}&       0.200\sym{***}&       0.200\sym{***}&       0.200\sym{***}&       0.200\sym{***}\\
		&     (0.011)         &     (0.011)         &     (0.011)         &     (0.011)         &     (0.011)         \\
		[1em]
			\end{tabular}
	\end{table}

\begin{table}[htbp]\footnotesize
\centering
%\caption{Benchmark models.}
\begin{tabular}{l*{5}{c}}
		November  &       0.358\sym{***}&       0.348\sym{***}&       0.348\sym{***}&       0.348\sym{***}&       0.348\sym{***}\\
		&     (0.011)         &     (0.011)         &     (0.011)         &     (0.011)         &     (0.011)         \\
		[1em]
		December  &       0.316\sym{***}&       0.306\sym{***}&       0.305\sym{***}&       0.306\sym{***}&       0.306\sym{***}\\
		&     (0.011)         &     (0.011)         &     (0.011)         &     (0.011)         &     (0.011)         \\
		\hline
		\(N\)       &      529,654         &      529,654         &      529,654         &      529,654         &      529,654         \\
		Adj. $R^2$&       0.657         &       0.659         &       0.659         &       0.659         &       0.659         \\ \hline\hline
		&&&&&\\
	
	%\begin{tablenotes}\\
	\multicolumn{6}{l}{\footnotesize{1. Standard errors are in parentheses}}\\
		\multicolumn{6}{l}{\footnotesize{2. ***, **, and * denote significance at the 1\%, 5\%, and 10\% levels, respectively.}}\\
		%2. \sym{*} \(p<0.05\), \sym{**} \(p<0.01\), \sym{***} \(p<0.001\)
		\multicolumn{6}{l}{\footnotesize{3. For artist, medium, auction house and auction city dummies, we only show}}\\
		\multicolumn{6}{l}{\footnotesize{ the results for the top ten categories ranked by their frequencies in the sample.}}
	%\end{tablenotes}
\end{tabular}
\end{table}

\newpage
\section{Topic Classification}
\label{appx c}
We use the first keyword(s) of the title to classify the works by topics. Both English and French keywords are included. We avoid using the keyword that can be used in different contexts.

These are the topic categories and their search strings: 

Abstract (``abstract", ``composition"),

Animals (``horse", ``cheval``, ``chevaux", ``cow\_", ``cows", ``vache", ``cattle", ``cat\_", ``cats", ``chat\_ ", ``dog\_", ``dogs", ``chien", ``sheep", ``mouton", ``bird", ``oiseau"), 

Landscape (``landscape", ``country landscape", ``coastal
landscape", ``paysage", ``seascape", ``sea\_", ``mer\_", ``mountain", ``river", ``riviere", ``lake", ``lac\_", ``valley", ``vallee"), 

Nude (``nude", ``nu\_", ``nue\_"),

People (``people", ``personnage", ``family", ``famille", ``boy", ``garcon", ``girl", ``fille", ``man\_", ``men\_", ``homme", ``woman", ``women", ``femme", ``child", ``enfant", ``couple", ``mother", ``mere\_",``father", ``pere\_", ``lady", ``dame"), 

Portrait (``portrait"),

Religion (``jesus", ``christ\_", ``apostle", ``ange\_", ``angel", ``saint\_", ``madonna", ``holy\_", ``mary magdalene", ``annunciation", ``annonciation", ``adoration", ``adam and
eve", ``adam et eve", ``crucifixion", ``last supper"), 

Self portrait (``self-portrait", ``self portrait", ``auto-portrait", ``autoportrait"), 

Still life (``still life", ``nature morte", ``bouquet"),

Urban(``city", ``ville", ``town", ``village", ``street", ``rue", ``market", ``marche", ``harbour", ``port\_", ``paris", ``london", ``londres", ``new york", ``amsterdam", ``rome\_", ``venice", ``venise").

Besides the above ten topic categories, we also define two additional ones: Untitled(``untitled", ``sans titre") and Unknown. For artworks that cannot be matched to any keyword, we put them into the Unknown category. Note that the sample of our robustness test with topic does not include the observations from these two categories (Untitled and Unknown).

\end{document}